\documentclass[twocolumn,superscriptaddress,preprintnumbers,amsmath,amssymb,prc]{revtex4-2}

\usepackage{graphicx}
\usepackage{dcolumn}
\usepackage{bm}
\usepackage{ifpdf}
\usepackage{epstopdf}
\usepackage{slashed}
\usepackage{amsfonts}
\usepackage{mathrsfs}
\usepackage{amssymb}
\usepackage{multirow}
\usepackage{CJK}
\usepackage{xcolor}
\usepackage{textcomp}
\usepackage{csquotes}

\begin{document}
\begin{CJK}{UTF8}{ipxm}
\preprint{RIKEN-iTHEMS-Report-22}

\title{Biexciton-like quartet condensates in an electron-hole liquid}

\author{Yixin Guo (郭一昕)}
\email{guoyixin1997@g.ecc.u-tokyo.ac.jp}
\affiliation{Department of Physics, Graduate School of Science, The University of Tokyo, Tokyo 113-0033, Japan}
\affiliation{RIKEN iTHEMS, Wako 351-0198, Japan}

\author{Hiroyuki Tajima (田島裕之)}
\email{htajima@g.ecc.u-tokyo.ac.jp}
\affiliation{Department of Physics, Graduate School of Science, The University of Tokyo, Tokyo 113-0033, Japan}

\author{Haozhao Liang (梁豪兆)}
\email{haozhao.liang@phys.s.u-tokyo.ac.jp}
\affiliation{Department of Physics, Graduate School of Science, The University of Tokyo, Tokyo 113-0033, Japan}
\affiliation{RIKEN Nishina Center, Wako 351-0198, Japan}

\date{\today}

\begin{abstract}
We theoretically study the ground-state properties and the condensations of exciton-like Cooper pairs and biexciton-like Cooper quartets in an electron-hole system.
Applying the quartet Bardeen-Cooper-Schrieffer (BCS) theory to the four-component fermionic system consisting of spin-$\frac{1}{2}$ electrons and spin-$\frac{1}{2}$ holes, we show how Cooper pairs and quartet correlations appear in the equation of state at the thermodynamic limit.
The biexciton-like four-body correlations survive even at the high-density regime as a many-body BCS-like state of Cooper quartets.
Our results are useful for further understanding of exotic matter in the interdisciplinary context of quantum many-body physics with multiple degrees of freedom.
\end{abstract}

\maketitle

\section{Introduction}\label{sec:I}

Quantum many-body systems exhibit nontrivial states which are absent in classical ones.
The interplay between quantum degeneracy and interactions leads to exotic condensation phenomena such as superfluidity and superconductivity~\cite{tilley2019superfluidity}.
The common states of matter surrounding us such as liquid droplets and crystalline solids are also deeply related to the interaction and quantum statistics of constituent particles from the microscopic viewpoint.

While it is known that superconductors and fermionic superfluids are triggered by the formation of two-body loosely bound states called Cooper pairs as a result of the Fermi-surface instability in the presence of two-body attractions~\cite{Bardeen1957Phys.Rev.108.11751204}, it is an interesting problem to explore condensation phenomena accompanying more than two-body bound states.
While spin-$\frac{1}{2}$ fermions with $s$-wave interaction tend to form two-body Cooper pairs because of their spin degree of freedom and Pauli's exclusion principles, multibody counterparts such as Cooper triples~\cite{Niemann2012Phys.Rev.A86.013628,Kirk2017Phys.Rev.A96.053614,Akagami2021Phys.Rev.A104.l041302,Tajima2021Phys.Rev.A104.053328} and quartets~\cite{Roepke1998Phys.Rev.Lett.80.31773180,Kamei2005J.Phys.Soc.Jpn.74.1911-1913,Sogo2010Phys.Rev.C81.064310,Senkov2011Phys.Atom.Nuclei74.12671276,Sandulescu2012Phys.Rev.C85.061303,Tsuruta2014J.Phys.Soc.Jpn.83.094603,Baran2020Phys.Lett.B805.135462,Baran2020Phys.Rev.C102.061301,guo2021cooper} can be formed in the presence of larger degrees of freedom for fermions (e.g., isospin, color, and atomic hyperfine states).

To study the nontrivial superfluid state associated with the Cooper instability leading to multibody bound states, semiconductor systems consisting of spin-$\frac{1}{2}$ electrons and holes can be promising candidates since these can be regarded as four-component fermionic systems with strong interactions.
In such systems, two- and four-body bound states called excitons and biexcitons are formed due to the attractive Coulomb electron-hole interaction~\cite{moskalenko2000bose}.
Moreover, the formation of polyexcitons consisting of more than two excitonic bound states was reported~\cite{Omachi2013Phys.Rev.Lett.111.026402}.
While the system is dominated by these bound states, e.g., excitons and biexcitons (or electron-hole plasma at finite temperature), in the low-carrier density regime, the quantum droplet appears as a many-body bound state in the higher density regime (before the semiconductor-metal transition) at low temperature~\cite{Combescot1972J.Phys.CSolidStatePhys.5.23692391,Brinkman1973Phys.Rev.B7.15081523,Comte1982J.Phys.43.10691081,Keldysh1986Contemp.Phys.27.395428}.
The Bardeen-Cooper-Schrieffer (BCS)-to-Bose-Einstein condensation (BEC) crossover associated with excitonic pairs with increasing the carrier density has been discussed extensively in previous theoretical works~\cite{Eagles1969Phys.Rev.186.456,Nozieres1985J.LowTemp.Phys.59.195211,Inagaki2002Phys.Rev.B65.205204,Pieri2007Phys.Rev.B75.113301,Ogawa2007J.Phys.Condens.Matter19.295205,Lozovik2008JETPLetters87.5559,Zenker2012Phys.Rev.B85.121102,Hanai2017Phys.Rev.B96.125206,Conti2017Phys.Rev.Lett.119.257002}. 
In highly-excited CuCl,
the condensation of biexcitons was observed~\cite{Chase1979Phys.Rev.Lett.42.12311234,Peyghambarian1983Phys.Rev.B27.23252345,Hasuo1993Phys.Rev.Lett.70.13031306}.
In the past years,
the formation of biexcitons was observed also in transition metal dichalcogenide crystals~\cite{You2015Nat.Phys.11.477481,Stevens2018Nat.Commun.9.,Steinhoff2018Nat.Phys.14.11991204,Chatterjee20212DMater.9.015023}.
Recently, it was reported that biexcitons play a key role for the formation of quantum droplets in photoexcited semiconductors~\cite{AlmandHunter2014Nature506.471475}.
Moreover, biexciton condensation has been found in an electron-hole Hubbard model at positive chemical potentials via a sign-problem-free quantum Monte Carlo simulation~\cite{Huang2020Phys.Rev.Lett.124.077601}.
Also, two-dimensional semiconductor systems in the biexciton-dominated regime have been investigated at finite temperature~\cite{Florez2020Phys.Rev.B102.115302}.
These studies suggest that it is important to clarify physical properties of the exciton and biexciton condensates for understanding many-body states at sufficiently low temperature.
 
Quartet condensation phenomena associated with the four-body bound states have also attracted much attention in nuclear systems~\cite{ring2004nuclear}.
Nuclear equations of state and their droplet properties are associated with strongly attractive nuclear forces leading to the formation of bound states such as deuterons, alpha particles, and heavier nuclei in the low-density region~\cite{Horowitz2006Nucl.Phys.A776.5579}, and the Fermi degenerate pressure of nucleons and multibody forces in the high-density region~\cite{Lattimer2012Annu.Rev.Nucl.Part.Sci.62.485515}.
Since alpha particles consisting of two neutrons and two protons is a stable cluster state with a large binding energy, the so-called alpha-particle condensation has been extensively studied in the context of Cooper quartets~\cite{Roepke1998Phys.Rev.Lett.80.31773180,Kamei2005J.Phys.Soc.Jpn.74.1911-1913,Sogo2010Phys.Rev.C81.064310,Senkov2011Phys.Atom.Nuclei74.12671276,Sandulescu2012Phys.Rev.C85.061303,Tsuruta2014J.Phys.Soc.Jpn.83.094603,Baran2020Phys.Lett.B805.135462,Baran2020Phys.Rev.C102.061301,guo2021cooper}.
Note that fluctuation-driven quartet formations have also been investigated in unconventional superconductors~\cite{Herland2010Phys.Rev.B82.134511,Grinenko2021Nat.Phys.17.12541259}.

Moreover, the quantum droplet state has been realized in ultracold Bose-Bose mixtures~\cite{Cabrera2018Science359.301304,Semeghini2018Phys.Rev.Lett.120.235301,Ferioli2019Phys.Rev.Lett.122.090401,DtextquotesingleErrico2019Phys.Rev.Research1.033155}.
The stabilization of the dilute quantum droplet is achieved by the competition between the mean-field attraction and the repulsive quantum fluctuations~\cite{Petrov2015Phys.Rev.Lett.115.155302}.
While the Lee-Huang-Yang energy density functional can explain such saturation properties but exhibit a complex value in the region where the mean-field collapse occurs,
it is reported that
the complexity of the energy density functional 
can be avoided by considering the bosonic pairing~\cite{Hu2020Phys.Rev.Lett.125.195302,Hu2020Phys.Rev.A102.043301}.
This fact implies that a biexciton, which can be regarded as the two-exciton pairing state, plays a crucial role in the formation of self-bound quantum droplets in electron-hole systems.
Moreover, similar self-bound quantum droplets have been realized in dipolar Bose gases~\cite{Schmitt2016Nature539.259262}, which is analogous with an exciton gas with an electric dipole moment.

In this paper, we theoretically investigate thermodynamic properties in an electron-hole system at zero temperature within the quartet BCS framework, which uses the extended BCS variational wave function involving Cooper pairing and quarteting in the momentum space at the thermodynamic limit~\cite{guo2021cooper}.  
Special attention is paid to the biexciton-like condensates, that is, the Cooper quartets consisting of two electrons and two holes as a result of the Cooper instability of Fermi seas.
(Note that we call it ``biexciton-like" since a Cooper quartet considered here is a loosely bound quantum state, unlike usual point-like bound states.)
Recently, such a framework has been employed to study pair and quartet correlations in nuclear systems~\cite{Baran2020Phys.Lett.B805.135462,Senkov2011Phys.Atom.Nuclei74.12671276,Baran2020Phys.Rev.C102.061301,guo2021cooper}.
Effects of Fermi degenerate pressure are automatically considered in this framework as in the usual BCS theory. 
The interplay among the Fermi degenerate pressure of electrons and holes and the formation of exciton-like Cooper pairs and biexciton-like Cooper quartets is examined microscopically.

This paper is organized as follows.
In Sec.~\ref{sec:II}, we show a theoretical model for an electron-hole system and a detailed formalism of the quartet BCS theory.
In Sec.~\ref{sec:III}, the numerical results and the corresponding discussions for the ground-state properties
are presented.
Finally, we summarize this paper with future perspectives in Sec.~\ref{sec:IV}.

\section{Theoretical Framework}\label{sec:II}

\subsection{Hamiltonian}\label{sec:IIA}

In this paper,  we consider a three-dimensional electron-hole system with the electron-electron, hole-hole, and electron-hole interactions.
The corresponding Hamiltonian is written as
\begin{align}
 H=
 H_{\rm{e}}^{0}
 + H_{\rm{h}}^{0}
 +V_{\rm{e-e}}
 +V_{\rm{h-h}}
 +V_{\rm{e-h}}.
\end{align}
In detail, the single-particle part reads
\begin{subequations}
\begin{align}
H_{\textrm{e}}^{0}=\,&
\sum_{\bm{p}, s_z} 
\varepsilon_{\textrm{e}, \bm{p}} 
e^\dagger_{\bm{p}, s_z}
e_{\bm{p}, s_z}
, \\
H_{\textrm{h}}^{0}=\,&
\sum_{\bm{p}, s_z} 
\varepsilon_{\textrm{h}, \bm{p}} 
h^\dagger_{\bm{p}, s_z}
h_{\bm{p}, s_z}, 
\end{align}
where the creation operators $e^\dagger$ and $h^\dagger$ create an electron and a hole, respectively;
$\bm{p}$ is the single-particle momentum, $\bm{q}=\frac{1}{2}(\bm{p}_1-\bm{p}_2)$ is the relative momentum, $s$ is the single-particle spin ($s_z$ is its third component), and $\bm{P}=\bm{p}_1+\bm{p}_2$ is the center of mass momentum. 
In addition, the single-particle energy reads
$\varepsilon_{i, \bm{p}}=
\frac{\bm{p}^{2}}{2 M_{\rm i}}-\mu_{{\rm i}}$ (${\rm i} =\textrm{e}, \textrm{h}$),
where $\mu_{{\rm i}}$ is the chemical potential and $M_{{\rm i}}$ is the effective mass.
Note that the particle-hole transformation is taken for the hole band such that a hole has the positive-curvature energy dispersion $\varepsilon_{\textrm{h},\bm{p}}$.
The low-energy interactions read
\begin{align}
\label{eq:int}
V_{\textrm{e-e}}=\,&
\sum_{\bm{P}, \bm{q}, \bm{q}'}
U_{\rm e-e}(\bm{q}-\bm{q}')
C^\dagger_{\textrm{e}}(\bm{P} ,\bm{q})
C_{\textrm{e}}(\bm{P} ,\bm{q}')
, \\
V_{\textrm{h-h}}=\,&
\sum_{\bm{P}, \bm{q}, \bm{q}'} 
U_{\rm h-h}(\bm{q}-\bm{q}')
C^\dagger_{\textrm{h}}(\bm{P} ,\bm{q})
C_{\textrm{h}}(\bm{P} ,\bm{q}')
,\\
V_{\textrm{e-h}}=\,&
{\frac{1}{4}}\sum_{S, S_z} 
\sum_{\bm{P}, \bm{q}, \bm{q}'} 
U_{\rm{e-h}}\left(\bm{q}-\bm{q}'\right) 
E^\dagger_{S, S_z}(\bm{P}, \bm{q})
E_{S, S_z}(\bm{P}, \bm{q}'),
\end{align}
\end{subequations}
where we have introduced the two-electron and two-hole pair operators
\begin{subequations}
\begin{align}
C^\dagger_{\textrm{e}}(\bm{P},\bm{q})=\,&
e^\dagger_{\bm{q}+\bm{P}/2, \frac{1}{2}}
e^\dagger_{-\bm{q}+\bm{P}/2, -\frac{1}{2}},\\
C^\dagger_{\textrm{h}}(\bm{P},\bm{q})=\,&
h^\dagger_{\bm{q}+\bm{P}/2, \frac{1}{2}}
h^\dagger_{-\bm{q}+\bm{P}/2, -\frac{1}{2}},
\end{align}
\end{subequations}
and the exciton creation operators
\begin{subequations}
\begin{align}
E^\dagger_{0, 0}(\bm{P}, \bm{q})
&=\sum_{s_z, s'_z} 
C_{\frac{1}{2} \frac{1}{2} s_z s'_z}^{00}
e^\dagger_{\bm{q}+\bm{P}/2, s_z} 
h^\dagger_{-\bm{q}+\bm{P}/2, s'_z},\\
E^\dagger_{1, S_z}(\bm{P}, \bm{q})
&=\sum_{s_z, s'_z} 
C_{\frac{1}{2} \frac{1}{2} s_z s'_z}^{1S_z}
e^\dagger_{\bm{q}+\bm{P}/2, s_z} 
h^\dagger_{-\bm{q}+\bm{P}/2, s'_z}.
\end{align}
\end{subequations}
Here, $S_z$ is the $z$ component of the total spin $S$ of an exciton.
The corresponding annihilation operators are their conjugates.
Also, $U_{\textrm{e-e}}$,
$U_{\textrm{h-h}}$
and $U_{\rm{e-h}}$ are the interaction strengths for the electron-electron, hole-hole, and electron-hole channels.
In general, the most relevant interaction is $U_{\rm e-h}$ which is an attractive Coulomb force and induces the formation of excitons.
For $U_{\textrm{e-e}}$ and $U_{\textrm{h-h}}$,
these can be attractive when the phonon-mediated interaction is present as in conventional BCS superconductors.
At high density, the Coulomb repulsion and the screening effect also may become important.
In this paper, we assume attractive $U_{\textrm{e-e}}$ and $U_{\textrm{h-h}}$ for simplicity
but eventually these interaction effects are ignored since attractive $U_{\rm e-h}$ is expected to be stronger than $U_{\textrm{e-e}}$ and $U_{\textrm{h-h}}$~\cite{Hanai2017Phys.Rev.B96.125206}. 

We briefly note that the present electron-hole system is like symmetric nuclear matter where the attractive electron-hole interaction can be regarded as a counterpart of the isospin-singlet neutron-proton interaction, which induces a two-body bound state (i.e., deuteron).
Indeed, both systems are composed of four-component fermions and similar multibody bound states appear in a certain density regime.
A simplified model enables us to discuss similarities and differences between two systems from an interdisciplinary viewpoint of many-body physics although their energy scales are largely different from each other.

\subsection{Quartet BCS theory}\label{sec:IIB}
With the consideration of the coherent state for the four-body sector, the trial wave function is adopted as~\cite{Senkov2011Phys.Atom.Nuclei74.12671276,guo2021cooper}
\begin{align}\label{twf2}
\left|\Psi\right\rangle=\,&
\prod_{\bm{q}}
\left[
u_{\bm{q}}
+
\frac{1}{2}\sum_{S, S_z} v_{\bm{q}, S, S_z} 
E_{S, S_z}^\dagger(0, \bm{q})
\right.\nonumber\\&\left.
+\sum_{i=\rm{e}, \rm{h}} x_{\bm{q}, i} C_{\rm i}^\dagger(\bm{0},\bm{q})
+\frac{1}{4}w_{\bm{q}} 
B^\dagger\left(\bm{q}\right)
\right]|0\rangle,
\end{align}
where the biexcition creation operator at the zero center-of-mass momentum is defined as
\begin{align}
B^\dagger\left(\bm{q}\right)=
E_{1, +1}^{\dagger}(0, \bm{q})
E_{1, -1}^{\dagger}(0, \bm{q}).
\end{align}
The contribution of excited excitons with finite center-of-mass momenta is neglected since the low-energy cluster states can dominate the system at sufficiently low temperatures.
We note that, a similar approximation has been employed in studies of nuclear systems~\cite{Senkov2011Phys.Atom.Nuclei74.12671276}.
The normalization condition is 
\begin{align}\label{normal}
\left|u_{\bm{q}}\right|^2+\left|\bm{v}_{\bm{q}}\right|^2+\left|\bm{x}_{\bm{q}}\right|^2+\left|w_{\bm{q}}\right|^2=1, 
\end{align}
where the norms of the variational parameters are defined as $\left|\bm{v}_{\bm{q}}\right|^2=\sum_{S, S_z}\left|v_{\bm{q}, S, S_z}\right|^2$ and $\left|\bm{x}_{\bm{q}}\right|^2=\sum_{\rm i}\left|x_{\bm{q}, \textrm{i}}\right|^2$ for convenience.

We note that while more sophisticated variational wave functions with the use of Hubbard-Stratonovich transformation are proposed in the studies of finite nuclei~\cite{Baran2020Phys.Lett.B805.135462,Baran2020Phys.Rev.C102.061301},
the present wave function has an advantage in the practical numerical calculation of the physical quantities at the thermodynamic limit because of its natural extension of the BCS wave function.  

The variational equations are obtained as 
\begin{widetext}
\begin{subequations}
\begin{align}
\label{eq:v}
v_{\bm{q}, 1, \pm1}=\,&
\frac{
 u_{\bm{q}}\Delta_{\bm{q}}^{\rm e-h}
 +
  w_{\bm{q}}{\Delta_{\bm{q}}^\ast}^{\rm e-h}}
{\Omega_{\bm{q}}+\left(\varepsilon_{\textrm{e}, \bm{q}}+\varepsilon_{\textrm{h}, -\bm{q}}\right)},
\qquad v_{\bm{q}, 1, 0}=
\frac{
 u_{\bm{q}}\Delta_{\bm{q}}^{\rm e-h}
 -
  w_{\bm{q}}{\Delta_{\bm{q}}^\ast}^{\rm e-h}}
{\Omega_{\bm{q}}+\left(\varepsilon_{\textrm{e}, \bm{q}}+\varepsilon_{\textrm{h}, -\bm{q}}\right)},
\qquad v_{\bm{q}, 0, 0}=
\frac{
 u_{\bm{q}}\Delta_{\bm{q}}^{\rm e-h}
 +
  w_{\bm{q}}{\Delta_{\bm{q}}^\ast}^{\rm e-h}}
{\Omega_{\bm{q}}+\left(\varepsilon_{\textrm{e}, \bm{q}}+\varepsilon_{\textrm{h}, -\bm{q}}\right)}.\\
x_{\bm{q}, \textrm{e}}=\,&\frac{
  u_{\bm{q}}\Delta_{\bm{q}}^{\rm e-e} }
{\Omega_{\bm{q}}+\left(\varepsilon_{\textrm{e}, \bm{q}}+\varepsilon_{\textrm{e}, -\bm{q}}\right)},
\qquad x_{\bm{q}, \textrm{h}}=\frac{
  u_{\bm{q}}\Delta_{\bm{q}}^{\rm h-h} }
{\Omega_{\bm{q}}+\left(\varepsilon_{\textrm{h}, \bm{q}}+\varepsilon_{\textrm{h}, -\bm{q}}\right)},\label{eq:x}\\
w_{\bm{q}}=\,&\frac{
v_{\bm{q}, 1, +1}\Delta_{\bm{q}}^{\rm e-h}
+v_{\bm{q}, 1, -1}\Delta_{\bm{q}}^{\rm e-h}
+v_{\bm{q}, 0, 0}\Delta_{\bm{q}}^{\rm e-h}
-v_{\bm{q}, 1, 0}\Delta_{\bm{q}}^{\rm e-h} }
{\Omega_{\bm{q}}+2\left(\varepsilon_{0, \bm{q}}+\varepsilon_{0, -\bm{q}}\right)},\label{eq:w}
\end{align}
\end{subequations}
where we introduced
\begin{align}
\label{eq:Omega}
\Omega_{\bm{q}}
=\,&
\frac{1}{2u_{\bm{q}}}
\left[ 
x^\ast_{\bm{q}, \textrm{e}}\Delta_{\bm{q}}^{\rm e-e}
+x_{\bm{q}, \textrm{e}}{\Delta^\ast_{\bm{q}}}^{\rm e-e} 
+x^\ast_{\bm{q}, \textrm{h}}\Delta_{\bm{q}}^{\rm h-h}
+x_{\bm{q}, \textrm{h}}{\Delta^\ast_{\bm{q}}}^{\rm h-h} 
+\sum_{S, S_z}v^\ast_{\bm{q}, S, S_z}\Delta_{\bm{q}}^{\rm e-h}
+\sum_{S, S_z}v_{\bm{q}, S, S_z}{\Delta^\ast_{\bm{q}}}^{\rm e-h}
\right].
\end{align}
The BCS-type energy gaps can be expressed in terms of the variational parameters as
\begin{subequations}
\begin{align}
\label{eq:gapee}
\Delta^{\rm{e-e}}_{\bm{q}}=\,&
-\sum_{\bm{q}'} 
U_{\rm e-e}(\bm{q}-\bm{q}')
u^\ast_{\bm{q}'}
x_{\bm{q}', \textrm{e}},\\
\Delta^{\rm{h-h}}_{\bm{q}}=\,&
-\sum_{\bm{q}'}  
U_{\rm h-h}(\bm{q}-\bm{q}')
u^\ast_{\bm{q}'}
x_{\bm{q}', \textrm{h}}, \label{eq:gaphh} \\
\Delta^{\rm{e-h}}_{\bm{q}}=\,&
-\sum_{\bm{q}'}
\sum_{S, S_z}
U_{\rm{e-h}}\left(\bm{q}-\bm{q}'\right) 
\left[
u^\ast_{\bm{q}'}
v_{\bm{q}', S, S_z}
+\delta_{S, 1}\delta_{S_z, +1}
v^\ast_{\bm{q}', S, -S_z}
w_{\bm{q}'}
+\delta_{S, 1}\delta_{S_z, -1}
v^\ast_{\bm{q}', S, -S_z}
w_{\bm{q}'}\right.\nonumber\\&\left.
-
\frac{1}{2}
\delta_{S, 1}
\delta_{S_z, 0}
\left(
v_{{\bm{q}'}, S, -S_z}^{\ast}
w_{\bm{q}'}
+
v_{{\bm{q}'}, S, -S_z}^{\ast}
w_{-\bm{q}'}
\right)
+
\frac{1}{2}
\delta_{S, 0}
\delta_{S_z, 0}
\left(
v_{{\bm{q}'}, S, -S_z}^{\ast}
w_{\bm{q}'}
+
v_{{\bm{q}'}, S, -S_z}^{\ast}
w_{-\bm{q}'}
\right)
\right]. \label{eq:gapeh}
\end{align}
\end{subequations}
\end{widetext}
The detailed derivations of the variational equations are further shown in Appendix~\ref{sec:IIC}.
In addition, we note that the well-known BCS results can be obtained by taking $w_{\bm{q}}=0$~\cite{guo2021cooper}.

To obtain the ground-state energy $E=\langle \Psi|H+\mu_{\rm e}n_{\rm e}+\mu_{\rm h}n_{\rm h}|\Psi\rangle$,
where
\begin{subequations}
\begin{align}
    n_{\rm e}&=\sum_{\bm{p},s_z}e_{\bm{p},s_z}^\dag e_{\bm{p},s_z},\\
    n_{\rm h}&=\sum_{\bm{p},s_z}h_{\bm{p},s_z}^\dag h_{\bm{p},s_z}
\end{align}
\end{subequations}
are the carrier density operators of electrons and holes, respectively,
we need to calculate the expectation values of $n_{\rm e}$ and $n_{\rm h}$.
These quantities (i.e., $\rho_{\rm e,h}=\langle\Psi|n_{\rm e,h}|\Psi\rangle$) are given by
\begin{subequations}
\begin{align}
\label{eq:rhoe}
\rho_{\rm e}=\,&\sum_{\bm{q}}\left(\left|\bm{v}_{\bm{q}}\right|^2+2\left|x_{\bm{q}, \textrm{e}}\right|^2+2\left|w_{\bm{q}}\right|^2\right),\\
\rho_{\rm h}=\,&\sum_{\bm{q}}\left(\left|\bm{v}_{\bm{q}}\right|^2+2\left|x_{\bm{q}, \textrm{h}}\right|^2+2\left|w_{\bm{q}}\right|^2\right).
\label{eq:rhoh}
\end{align}
\end{subequations}

In the numerical calculations,
we solve Eqs.~(\ref{eq:gapee}), (\ref{eq:gaphh}), (\ref{eq:gapeh}), (\ref{eq:Omega}), (\ref{eq:v}), (\ref{eq:x}), and (\ref{eq:w}) with respect to $\Delta_{\bm{q}}^{\rm e-e}$, $\Delta_{\bm{q}}^{\rm h-h}$, $\Delta_{\bm{q}}^{\rm e-h}$, $\Omega_{\bm{q}}$, $v_{\bm{q}}$, $x_{\bm{q}}$, and $w_{\bm{q}}$ self-consistently.
Then, $u_{\bm{q}}$ is determined by the normalization condition in Eq.~(\ref{normal}).
Substituting these variational parameters to Eqs.~(\ref{expH}), (\ref{eq:rhoe}), and (\ref{eq:rhoh}),
we can numerically evaluate the ground-state energy $E=\langle H\rangle +\mu_{\rm e}\rho_{\rm e}+\mu_{\rm h}\rho_{\rm h}$ and the fermion number density $\rho=\rho_{\rm e}+\rho_{\rm h}$. 

Practically, in this paper we consider only the short-range attractive electron-hole interaction described by the contact-type coupling as $U_{\rm e-h}(\bm{q}-\bm{q}')=-U_{\rm C}$~\cite{Hanai2017Phys.Rev.B96.125206}.
A similar contact-type coupling has also been employed for the study of monolayer MoSe$_2$ \cite{Kwong2021PhysRevB.104.245434}.
Also, we consider the equal effective masses as $M_{\rm e}=M_{\rm h}\equiv M$.
Although it is rather simplified compared with the realistic cases, such a model is sufficient for our purpose since we are interested in qualitative features of BCS-like pair and quartet correlations in an electron-hole system.
Indeed, the long-range Coulomb attraction is necessary to be considered for the description of the droplet state~\cite{Combescot1972J.Phys.CSolidStatePhys.5.23692391,Brinkman1973Phys.Rev.B7.15081523,Comte1982J.Phys.43.10691081,Keldysh1986Contemp.Phys.27.395428}.
Nevertheless, our approach is useful for understanding the Cooper pair and quartet correlations on the ground-state energy.

\section{Results and Discussion}\label{sec:III}

To figure out the differences between the results with and without the biexciton-like Cooper quartet correlations, we take the electron $M_e$ and hole mass $M_h$ to be the same as $0.511$~MeV, the four-body (biexciton) energy $B_{\rm XX}$ as $500$~meV to characterize the electron-hole interaction strength $U_{\rm C}$, and the momentum cutoff $\Lambda=100k_{\rm F}$, where $k_{\rm F}=(3\pi^2\rho/2)^{\frac{1}{3}}$ is the Fermi momentum.
It should be noted that $B_{\rm XX}= 500$ meV is close to the value $434$ meV employed in Ref.~\cite{Florez2020Phys.Rev.B102.115302}.

\subsection{Ground-state properties of biexciton-like quartet condensates in an electron-hole system}\label{sec:IIIA}

\begin{figure}[t]
  \includegraphics[width=0.45\textwidth]{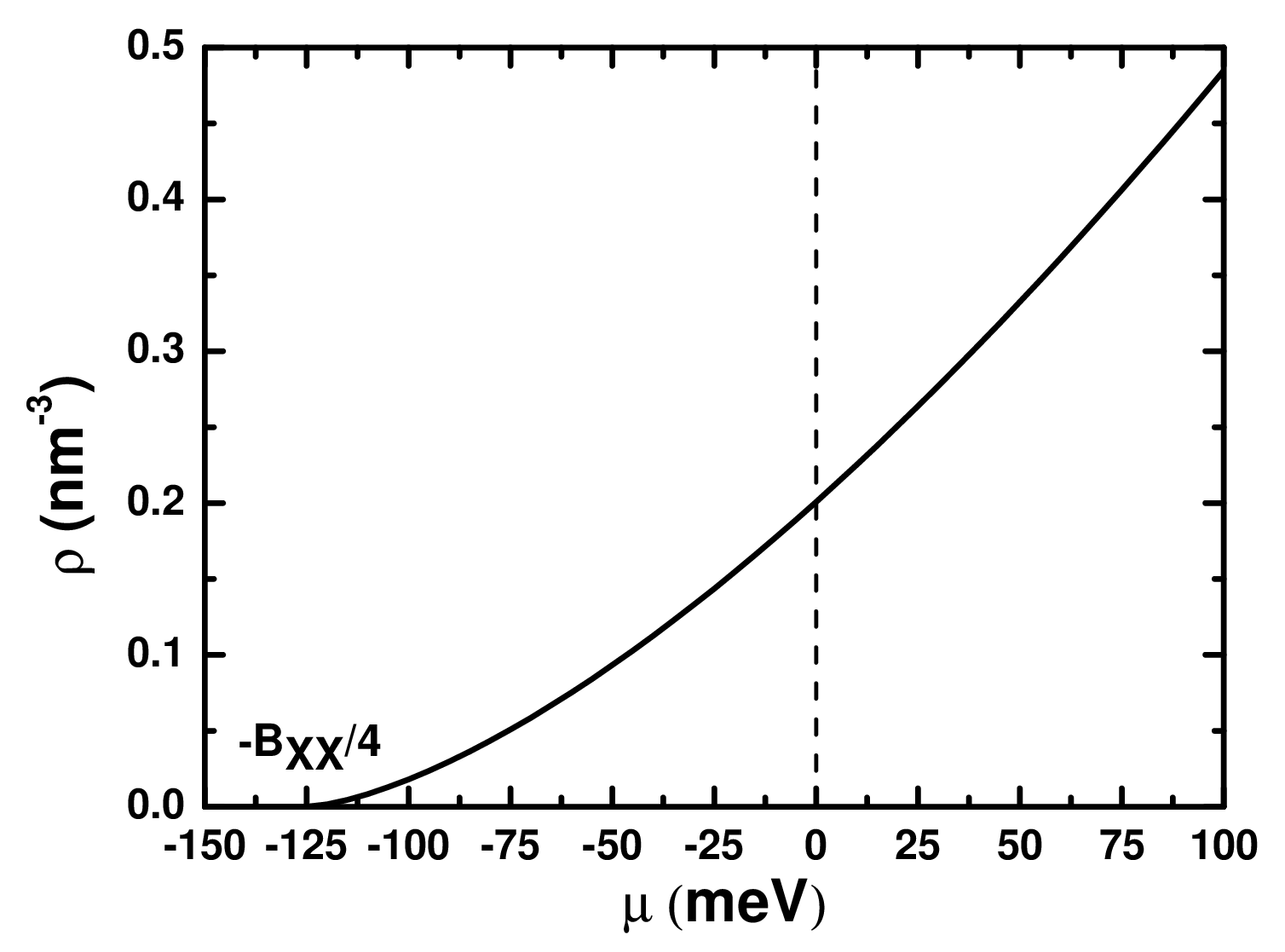}
  \caption{Total fermion number density $\rho$ as a function of chemical potential $\mu$. The four-body energy $B_{\rm XX}$ corresponds to the point where the total density becomes finite.
  Here, the four-body energy per fermion $B_{\rm XX}/4$ is taken as $125$~meV.
  }\label{fig:1}
\end{figure}

In the low-density limit, the ground-state energy density $E$ is proportional to the cluster energy as~\cite{Navon2010Science328.729732}
\begin{align}
\label{eq:elowd}
    E_{\rho\rightarrow 0}=-\frac{1}{4}B_{\rm XX}\rho.
\end{align}
Since the fermion chemical potential $\mu_{\rm e}=\mu_{\rm h}\equiv \mu$ for the balanced system ($\rho_{\rm e}=\rho_{\rm h}$ and $M_{\rm e}=M_{\rm h}$) is given by $\mu=\left(\frac{\partial E}{\partial \rho}\right)$ based on the thermodynamic relation,
one can obtain $-B_{\rm XX}=4\mu~(\rho\rightarrow 0)$.
Figure~\ref{fig:1} shows the total fermion number density $\rho$ as a function of $\mu$ for the electron-hole interaction strength $U_{\rm C}$ that corresponds to $B_{\rm XX} = 500$~meV. Note that $B_{\rm XX}$ is associated with $U_{\rm C}$ through the variational equations~(\ref{eq:v}) and (\ref{eq:w}) and electron-hole pairing gap given by Eq.~(\ref{eq:gapeh}), so that the value of $B_{\rm XX}$ varies if $U_{\rm C}$ changes and vice versa.
In this figure, it is clearly seen that $\rho$ starts to be finite at $\mu = - B_{\rm XX}/4 = -125$~meV.

For the two-body sector, because the two-body (exciton) energy $B_{\rm X}$ cannot be determined from $E$, we evaluate $B_{\rm X}$ by solving the two-body problem with the same $U_{\rm C}$.
The relation between $U_{\rm C}$ and $B_{\rm X}$ is summarized in Appendix~\ref{appendix}.
We numerically confirmed that $B_{\rm XX}$ is larger than $2B_{\rm X}$ in the region where we explored in the present model.
While it is difficult to prove this relation of $B_{\rm XX}$ and $B_{\rm X}$ for arbitrary coupling strength,
our trial wave function can describe both pair and quartet states in the common variational parameter space.
Therefore, based on the variational principle, it indicates that the biexciton state is stable against the breakup to two exciton states in the dilute limit.
It is known that for the contact-type interaction the cutoff dependence will appear in the numerical calculations, and a density-dependent cutoff is adopted here. 
However, we calculate $B_{\rm X}$ according to Appendix~\ref{appendix} in the low-density limit ($\rho\sim10^{-6}$~nm$^{-3}$) and obtain that $B_{\rm X}\simeq 225$ meV.
Consequently, we regard that the two-body energy $B_{\rm X}= 225$ meV in vacuum.
It is close to the value of exciton energy, $193$ meV, adopted in Ref.~\cite{Florez2020Phys.Rev.B102.115302}.
Note that if we measure the biexciton binding energy $E_{\rm XX}$ from the threshold for two exciton states given by $2E_{\rm X}=-2B_{\rm X}$, we obtain $E_{\rm XX}=-B_{\rm XX}-2{E_{\rm X}}=-50$ meV, which is also close to $-43$ meV in Ref.~\cite{Florez2020Phys.Rev.B102.115302}.
In addition, although the calculations performed in this paper are basically for the three-dimensional system, the present theoretical framework can be further applied to the two-dimensional ones by taking $D=2$ in the momentum summation, $\sum_{\bm{q}}\rightarrow\int\frac{d^D\bm{q}}{(2\pi)^D}$.
For instance, our framework is closely related to the model for CdSe nanoplatelets in Ref.~\cite{Florez2020Phys.Rev.B102.115302} with a different dimension.
Another relevant study
~\cite{Conti2020Phys.Rev.B101.220504} was performed in the two-dimensional van der Waals materials with a long-range (momentum-dependent) interaction, where the coupled MoSe$_2$-WSe$_2$ monolayers were taken as the objects of research.
Nevertheless, the biexciton-like quartet correlation was not taken into account in those works.
Therefore, the present theoretical framework can be applied to more realistic systems.

\begin{figure}[t]
  \includegraphics[width=0.45\textwidth]{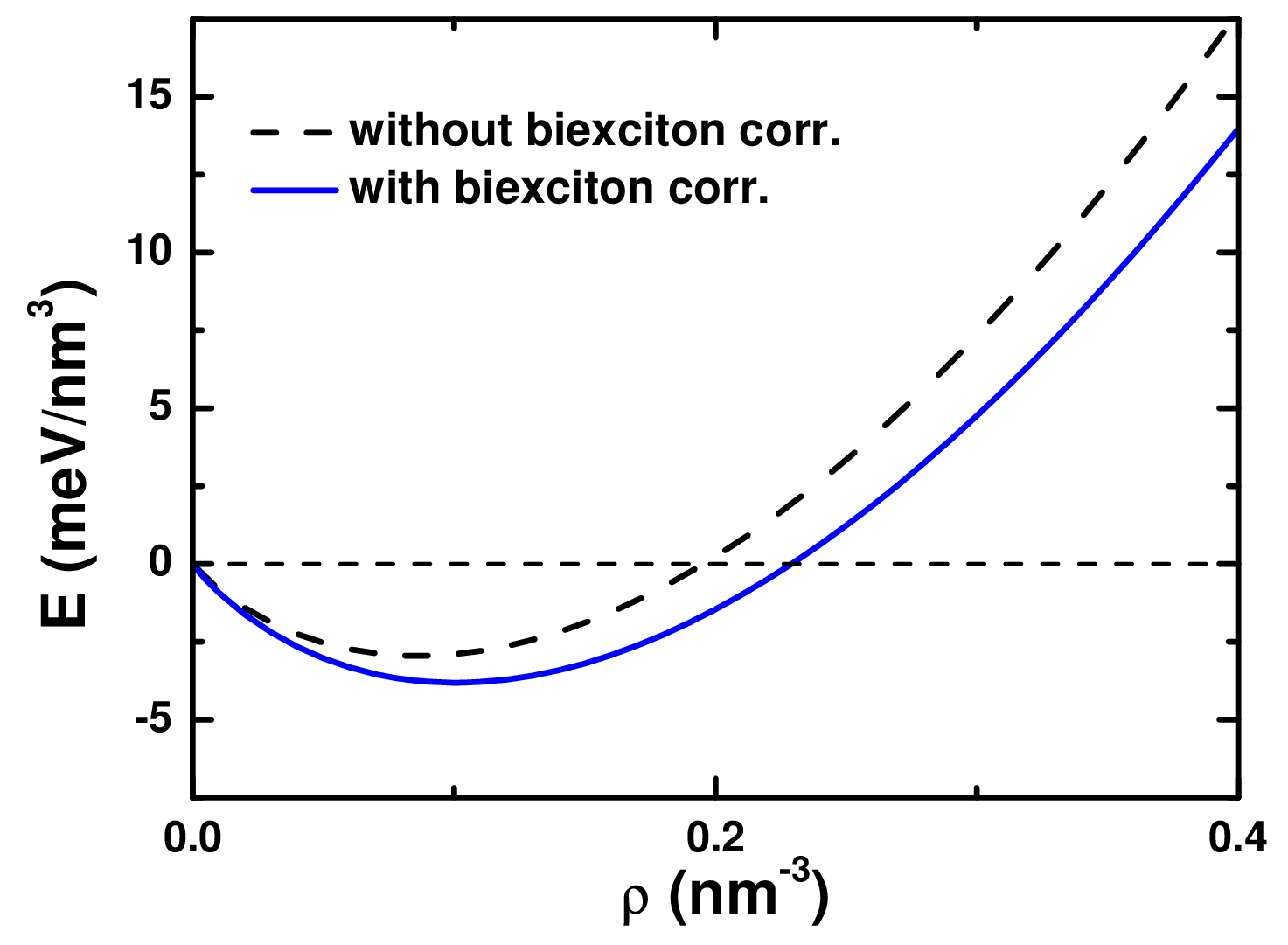}
  \caption{(Color online) Energy densities $E$ as a function of total density energy $\rho$ with (blue solid line) and without (black dashed line) biexciton correlation. 
  The four-body energy $B_{\rm XX}$ is taken as $500$~meV.}\label{fig:2}
\end{figure}

Figure~\ref{fig:2} shows the ground-state energy density $E=\langle H\rangle+\mu \rho$ as a function of $\rho$.
To see the role of quartet correlations, the energy density $E$ without biexciton correlation is also plotted.
Because the bound-state formation reduces the total energy,
the equation of state becomes softer (i.e., the ground-state energy becomes smaller) compared with the result without biexciton correlations.
As shown in Eq.~(\ref{eq:elowd}), $E$ decreases with increasing $\rho$ in the low-density regime,
indicating that the system obtains the energy gain associated with the bound-state formations (i.e., excitons and biexcitons).
In turn, the absolute value of the quartet correction, indicated by the difference between the results with and without biexciton correlation, becomes larger with increasing $\rho$.
This result indicates that the Cooper instability associated with the Fermi surface and the attractive electron-hole interaction assists the formation of Cooper quartets in the high-density regime.
In this sense, the in-medium biexciton correlations in such a dense system are not the usual four-body bound states in vacuum but the BCS-like many-body states of biexciton-like Cooper quartets,
which are also different from polyexcitons.

In the quartet BCS framework, the low-energy excitation is dominated by the quartet correlations.
In the high-density regime, such a low-energy sector relatively increases with the increase of the Fermi energy.
However, the quartet correlations themselves are negligible compared with the Fermi energy in such a regime.
Although we do not explicitly show it here, the increase of $E$ in the high-density regime can be understood from the behavior of the energy density $E_{\rm FG}$ in an ideal Fermi gas
\begin{align}
    E_{\rm FG}=\frac{2}{5\pi^2M}\left(\frac{3\pi^2\rho}{2}\right)^{\frac{5}{3}},
\end{align}
which is a monotonically increasing function with respect to $\rho$.
We note that a triexciton, which is a six-body bound state consisting of three electrons and three holes,
is not considered in this paper
because the Pauli-blocking effect tends to suppress such bound states involving more than two fermions with the same spins for the $s$-wave short-range interactions.

While the disappearance of quartet correlations with increasing density was reported in nuclear matter~\cite{Roepke1998Phys.Rev.Lett.80.31773180,Sogo2010Phys.Rev.C81.064310}, it is deeply related to the form of the two-body interaction, such as the effective range corrections and the higher-partial waves, as well as the three- and four-body interactions.
Since we employ the contact-type two-body coupling with a large momentum cutoff $\Lambda=100k_{\rm F}$,
pair and quartet correlations are not suppressed in the high-density regime explored in this study.
This result is also associated with the fact that the high-density regime in our model with a contact coupling does not correspond to the usual weak-coupling case as in conventional BCS superconductors but rather the unitary (or crossover) regime from the viewpoint of the BCS-BEC crossover because $U_{\rm C}$ involves the two-body bound state (i.e., positive scattering length) in the free space~\cite{Andrenacci1999Phys.Rev.B60.1241012418}.
On the other hand, at finite temperature, the phase transition from Cooper-quartet condensates to an electron-hole plasma may occur even in the present model.
More detailed investigations with realistic interactions in the high-density regime and the semiconductor-metal transition are
out of scope of this paper and will be addressed elsewhere.

Moreover, we do not find a minimum of $E/\rho$ (i.e., the energy per one fermion) with respect to $\rho$, implying the absence of the droplet phase due to the artifact of the contact-type interactions in the present model.
To overcome this, we need to consider the finite-range attractive interaction giving a finite Hartree-Fock contribution, which is approximately proportional to $-\rho^2$~\cite{kadanoff2018quantum}.
Nevertheless, the present results showing how the quartet correlations affect the energy density could be useful for future detailed investigations of droplet phase with more realistic interactions.

\subsection{Energy dispersion and excitation gap}

In this subsection, we discuss how the quartet correlations affect the excitation energy of the system.
First, in the absence of quartet correlations ($w_{\bm{q} }= 0$),
one can obtain
\begin{align}
   \Omega_{\bm{q}}=E_{\bm{q}}-\varepsilon_{\bm{q}}, 
\end{align}
where
\begin{align}\label{eq:E_BCS}
E_{\bm{q}}=\sqrt{\varepsilon_{\bm{q}}^{2}+\bm{\Delta}_{\bm{q}}^{2}}
\end{align}
is the usual BCS dispersion with $\bm{\Delta}_{\bm{q}}^2=\sum_{S,S_z}|\Delta_{\bm{q}}^{\rm e-h}|^2$.
One can obtain the excitation gap $E_{\rm gap}={\rm inf}_{\bm{q}}[2E_{\bm{q}}]\equiv 2|\bm{\Delta}_{\bm{q}=\bm{q}_{\rm min}}|$, where $\bm{q}_{\rm min}$ is the momentum on the bottom of $E_{\bm{q}}$.
Note that $|\bm{q}_{\rm min}|=\sqrt{2M\mu}$ in the present case with the contact coupling.
In the presence of quartet correlations (i.e., $\omega_{\bm{q}}\neq 0$),
one can obtain
\begin{align}
\label{eq:qOmega}
    \Omega_{\bm{q}}=E_{\bm{q}}^{\omega}-\varepsilon_{\bm{q}},
\end{align}
where
\begin{align}
\label{eq:E_qBCS}
    E_{\bm{q}}^{\omega}=\sqrt{E_{\bm{q}}^2+\frac{4|\Delta_{\bm{q}}^{\rm e-h}|^4}{(\Omega_{\bm{q}}+2\varepsilon_{\bm{q}})(\Omega_{\bm{q}}+4\varepsilon_{\bm{q}})-4|\Delta_{\bm{q}}^{\rm e-h}|^2}}.
\end{align}
In analogy with the usual BCS dispersion (\ref{eq:E_BCS}),
$E_{\bm{q}}^{\omega}$ can be regarded as the quartet BCS dispersion~\cite{guo2021cooper}. 
Solving Eq.~(\ref{eq:E_qBCS}) combined with Eq.~(\ref{eq:qOmega}),
one can evaluate the excitation gap $E_{\rm gap}={\rm inf}_{\bm{q}}[2E_{\bm{q}}^{\omega}]$ in the quartet BCS framework.

\begin{figure}[t]
  \includegraphics[width=0.45\textwidth]{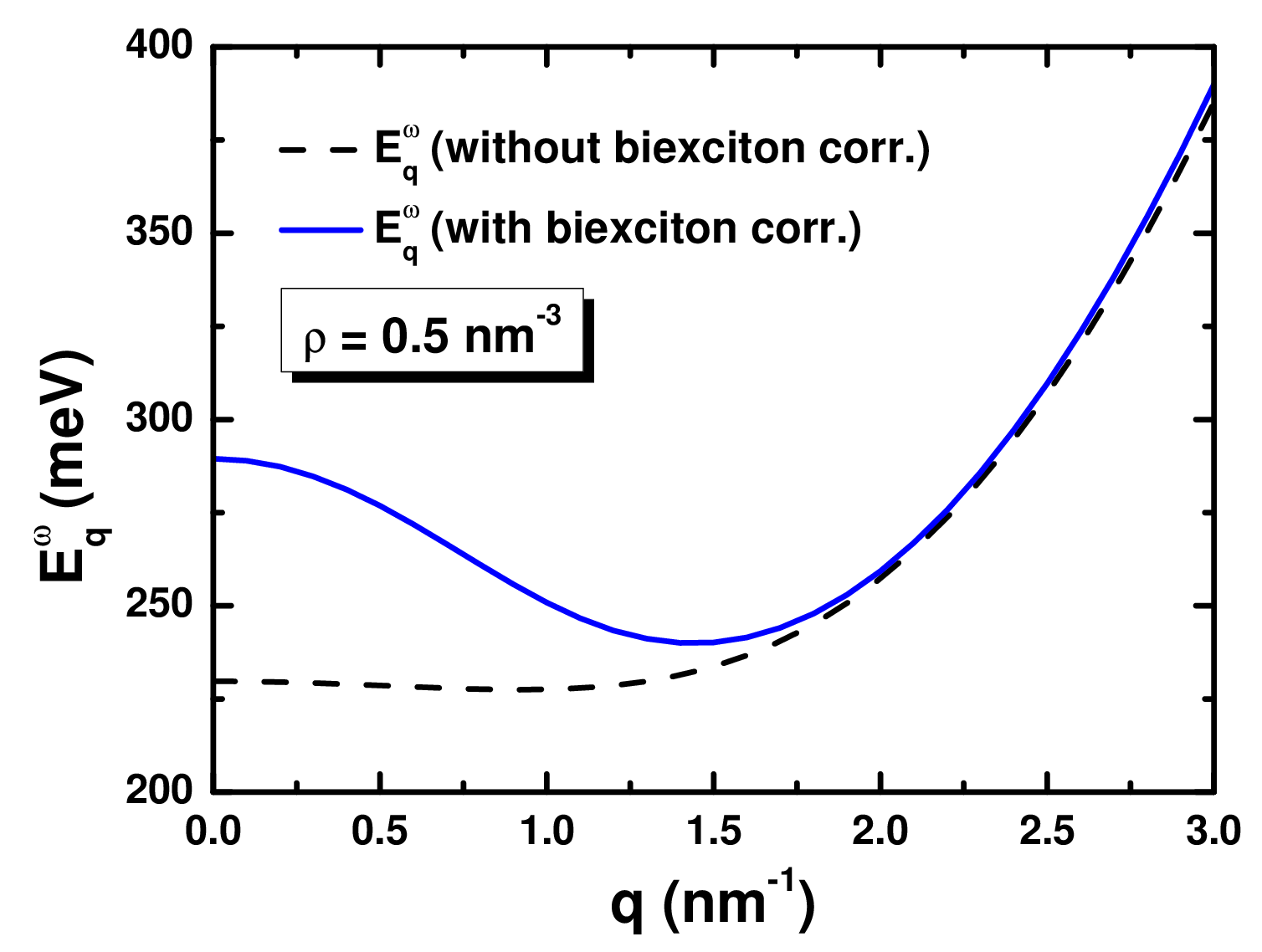}
  \caption{(Color online) Energy dispersion $E_{\bm{q}}^{\omega}$ for a given density $\rho=0.5$~nm$^{-3}$ with (blue solid line) and without (black dashed line) biexciton-like quartet correlations as a function of relative momentum $\bm{q}$.
  }\label{fig:3}
\end{figure}

The energy dispersions with and without the biexciton correlations (i.e., $E_{\bm{q}}^{\omega}$ and $E_{\bm{q}}$) as a function of relative momentum $q=|\bm{q}|$ are shown in Fig.~\ref{fig:3}, where we take $B_{\rm XX}=500$ meV.
Because we are interested in the quartet BCS regime where $\mu$ becomes positive and the Fermi surface effect is important,
the high density case with $\rho=0.5$ nm$^{-3}$ is examined here.
As shown in Fig.~\ref{fig:1}, $\mu$ reaches $100$ meV at $\rho=0.5$ nm$^{-3}$.
With the consideration of biexciton correlations, the excitation gap $E_{\rm gap}$, namely, the minimum of the energy dispersion, becomes larger by around $5.5\%$, and the relative momentum which gives the minimum of energy dispersion also becomes larger by around $55.6\%$.
While the quartet corrections are significant in the low-momentum regime, $E_{\bm{q}}^\omega$ becomes closer to $E_{\bm{q}}$ in the high-momentum regime.
Thus, one can see that $E_{\bm{q}}^\omega$ increases at low $\bm{q}$ compared with $E_{\bm{q}}$ because of the quartet corrections as found in Eq.~(\ref{eq:E_qBCS}). 
This result indicates that excitons consisting lower relative momenta tends to form the biexciton-like Cooper quartets and such quartets are energetically broken into two exciton-like Cooper pairs for larger $\bm{q}$. 

\begin{figure}[t]
  \includegraphics[width=0.45\textwidth]{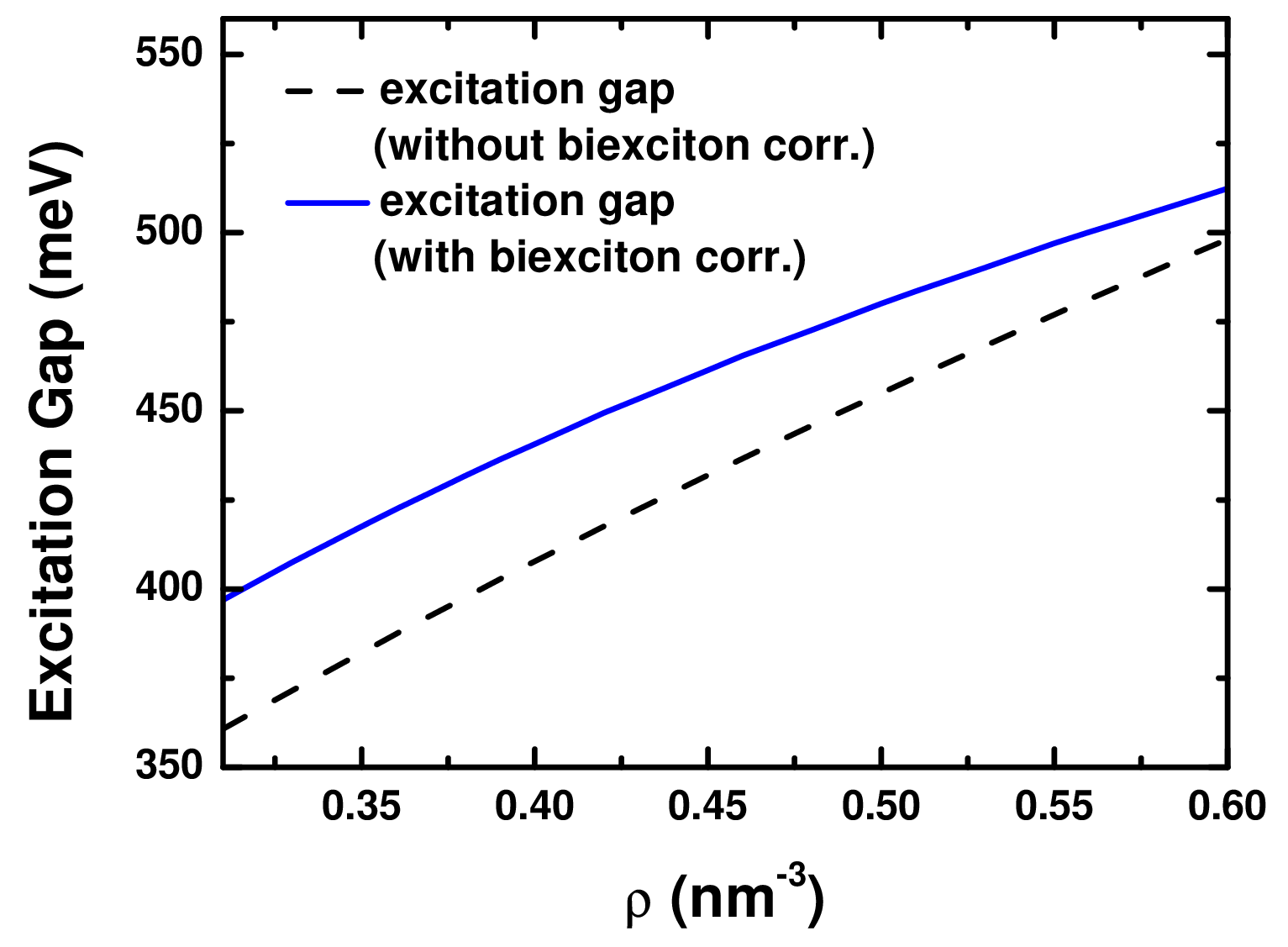}
  \caption{(Color online) Excitation gap $E_{\rm gap}$ with (blue solid line) and without (black dashed line) biexciton-like quartet correlations as a function of total density $\rho$.
  }\label{fig:4}
\end{figure}

Finally, in Fig.~\ref{fig:4}, we plot $E_{\rm gap}$ with quartet correlations estimated from the minimum of $E_{\bm{q}}^\omega$ shown in Fig.~\ref{fig:3}.
For comparison, we also show the result of the excitation gap without quartet correlations.
In general, $E_{\rm gap}$ with quartet correlations becomes larger than the case without them.
This behavior is natural since a larger energy is needed to excite a single carrier accompanying the breakup of quartets compared with the case with only two-body pairings.
Also, one can find that the difference between the cases with and without quartet correlations becomes smaller with increasing $\rho$.
At first glance, this tendency seems to be opposite to the quartet correlations on the ground-state energy $E$ shown in Fig.~\ref{fig:3}, but actually these results are found to be consistent by considering how these quantities are associated with quartet correlations in a relative-momentum-resolved way.
While the lower relative-momentum sector plays a significant role for the quartet corrections on $E$ involving the $\bm{q}$ summation,
$E_{\rm gap}$ reflects the quartet correlations at $\bm{q}=\bm{q}_{\rm min}$, which is relatively large compared with the low relative momenta dominated by the quartet formation. 
Indeed, the difference between $E_{\bm{q}}^{\omega}$ and $E_{\bm{q}}$ near $\bm{q}=\bm{q}_{\rm min}$ is smaller compared with that at $\bm{q}\simeq \bm{0}$.
In this regard, spectroscopic measurements for in-medium biexciton energy, which are not momentum-resolved, would give the similar tendency of $\rho$ dependence as shown in Fig.~\ref{fig:4}.

\section{Summary and Perspectives}\label{sec:IV}

In this paper, we investigated the microscopic properties of biexciton-like quartet condensates in an electron-hole system within the quartet BCS theory at the thermodynamic limit.
The variational approach is applied to the three-dimensional electron-hole system, which is described as four-component fermions with short-range attractive interactions (corresponding to the Coulomb electron-hole attraction).
Numerically solving the variational equations, we have obtained the ground-state energy density as a function of the fermion number density.

On the one hand, the ground-state energy density decreases with increasing number density in the dilute region because of the energy gains associated with the biexciton formations.
On the other hand, such a tendency for the ground-state energy density turns into the increase in the high-density regime due to the Fermi degenerate pressure.
To see the role of quartet correlations,
we compared the results with and without quartet correlations and pointed out that the quartet condensation leads to the lower ground-state energy.
Moreover, we showed the density dependence of the excitation gap, which is defined as the minimal dispersion in analogy with the usual BCS theory.
While the quartet correlations induce a larger excitation gap in the whole density regime,
the difference from the result with only pairing correlations can be smaller in the high-density regime, because the dispersion minimum itself does not involve the quartet correlations associated with lower momenta.

In this paper, we have employed a simplified model to explore qualitative features of the condensation energy of the biexciton-like quartet state.
For further quantitative investigations of the electron-hole droplet phase,
it is needed to apply more realistic models with long-range interactions (e.g., Coulomb interactions and their screening) and multi-body forces.
For semiconductor systems such as layered transition metal dichalcogenides,
the two-dimensional model is relevant.
While the quadratic dispersion is adopted in this paper,
the band structure of each material should be considered.
Nevertheless, these extensions can be easily done in our quartet BCS theory at the thermodynamic limit.
It can be achieved by replacing
the three-dimensional momentum summation with a two-dimensional one and 
fermionic dispersion $\varepsilon_{i,\bm{p}}$ with realistic bands, respectively.

Also, quantum fluctuations associated with excited two- and four-body states can be important.
The energy density functional involving these corrections would be useful for further developments not only in condensed matter but also nuclear and cold atomic physics.
Moreover, since actual electron-hole systems are realized as a non-equilibrium steady state,
the interactions with environments as an open quantum system would also be an interesting topic.
These are left for future works.

\begin{acknowledgments}

Y.G. was supported by the RIKEN Junior Research Associate Program.
H.T. acknowledges the JSPS Grants-in-Aid for Scientific Research under Grant No.~18H05406.
H.L. acknowledges the JSPS Grant-in-Aid for Early-Career Scientists under Grant No.~18K13549, the JSPS Grant-in-Aid for Scientific Research (S) under Grant No.~20H05648, and the RIKEN Pioneering Project: Evolution of Matter in the Universe.

\end{acknowledgments}

\appendix

\section{The derivation of the variational equations}\label{sec:IIC}
In this appendix, we derive the variational equation for biexciton-like quartet condensates.

\begin{widetext}

The expectation value of the Hamiltonian is evaluated as
\begin{align}
\label{expH}
\langle \Psi \left|H\right| \Psi\rangle
=\,&\langle \Psi \left|H_0\right| \Psi\rangle
+\langle \Psi \left|V_{\rm{e-e}}\right| \Psi\rangle
+\langle \Psi \left|V_{\rm{h-h}}\right| \Psi\rangle
+\langle \Psi \left|V_{\rm{e-h}}\right| \Psi\rangle
\nonumber\\
=\,&\sum_{\bm{q}}\left[
\left(
\left|v_{\bm{q}, 1, +1}\right|^2
+\left|v_{\bm{q}, 1, -1}\right|^2
+\left|v_{\bm{q}, 1, 0}\right|^2
+\left|v_{\bm{q}, 0, 0}\right|^2
\right)
\left(
\varepsilon_{\textrm{e}, \bm{q}}
+\varepsilon_{\textrm{h}, -\bm{q}}
\right)
\right.\nonumber\\
&\left.+\left|x_{\bm{q}, \textrm{e}}\right|^2
\left(
\varepsilon_{\textrm{e}, \bm{q}}
+\varepsilon_{\textrm{e}, -\bm{q}}
\right)
+\left|x_{\bm{q}, \textrm{h}}\right|^2
\left(
\varepsilon_{\textrm{h}, \bm{q}}
+\varepsilon_{\textrm{h}, -\bm{q}}
\right)\right.\nonumber\\
&\left.
+2\left|w_{\bm{q}}\right|^2\left(
\varepsilon_{0, \bm{q}}
+\varepsilon_{0, -\bm{q}}
\right)
\right]\nonumber\\
&+ 
\sum_{\bm{q}, \bm{q}'}  
U_{\rm e-e}(\bm{q}-\bm{q}')\left(
x^\ast_{\bm{q}, \textrm{e}}
u_{\bm{q}}
\right)\cdot
\left(
u^\ast_{\bm{q}'}
x_{\bm{q}', \textrm{e}}
\right) 
+\sum_{\bm{q}, \bm{q}'}  
U_{\rm h-h}(\bm{q}-\bm{q}')\left(
x^\ast_{\bm{q}, \textrm{h}}
u_{\bm{q}}
\right)\cdot
\left(
u^\ast_{\bm{q}'}
x_{\bm{q}', \textrm{h}}
\right)\nonumber\\
&+
\sum_{\bm{q}, \bm{q}'}\sum_{S, S_z}
U_{\rm{e-h}}\left(\bm{q}-\bm{q}'\right) 
\left[
u_{\bm{q}}
v^\ast_{\bm{q}, S, S_z}
+\delta_{S, 1}\delta_{S_z, +1}
v_{\bm{q}', S, -S_z}
w^\ast_{\bm{q}'}
+\delta_{S, 1}\delta_{S_z, -1}
v_{\bm{q}', S, -S_z}
w^\ast_{\bm{q}'}\right.\nonumber\\&\left.
-
\frac{1}{2}
\delta_{S, 1}
\delta_{S_z, 0}
\left(
v_{{\bm{q}'}, S, -S_z}
w_{\bm{q}'}^{\ast}
+
v_{-{\bm{q}'}, S, -S_z}
w_{\bm{q}'}^{\ast}
\right)
+
\frac{1}{2}
\delta_{S, 0}
\delta_{S_z, 0}
\left(
v_{{\bm{q}'}, S, -S_z}
w_{\bm{q}'}^{\ast}
+
v_{-{\bm{q}'}, S, -S_z}
w_{\bm{q}'}^{\ast}
\right)
\right]
\nonumber\\
&\times
\left[
u^\ast_{\bm{q}'}
v_{\bm{q}', S, S_z}
+\delta_{S, 1}\delta_{S_z, +1}
v^\ast_{\bm{q}', S, -S_z}
w_{\bm{q}'}
+\delta_{S, 1}\delta_{S_z, -1}
v^\ast_{\bm{q}', S, -S_z}
w_{\bm{q}'}\right.\nonumber\\&\left.
-
\frac{1}{2}
\delta_{S, 1}
\delta_{S_z, 0}
\left(
v_{{\bm{q}'}, S, -S_z}^{\ast}
w_{\bm{q}'}
+
v_{{\bm{q}'}, S, -S_z}^{\ast}
w_{-\bm{q}'}
\right)
+
\frac{1}{2}
\delta_{S, 0}
\delta_{S_z, 0}
\left(
v_{{\bm{q}'}, S, -S_z}^{\ast}
w_{\bm{q}'}
+
v_{{\bm{q}'}, S, -S_z}^{\ast}
w_{-\bm{q}'}
\right)
\right].
\end{align}
By taking the variations of the expectation value of the Hamiltonian with respect to variational parameters, we obtain
\begin{align}
&\delta\langle \Psi \left|H\right| \Psi\rangle\nonumber\\
=\,&
\sum_{S, S_z}
v_{\bm{q}, S, S_z}\delta v^\ast_{\bm{q}, S, S_z}
\left(
\varepsilon_{\textrm{e}, \bm{q}}
+\varepsilon_{\textrm{h}, -\bm{q}}
\right)
+\sum_{\rm{i}}x_{\bm{q}, \textrm{i}}\delta x^\ast_{\bm{q}, \textrm{i}}
\left(
\varepsilon_{\textrm{i}, \bm{q}}
+\varepsilon_{\textrm{i}, -\bm{q}}
\right)
+2w_{\bm{q}}\delta w^\ast_{\bm{q}}\left(
\varepsilon_{0, \bm{q}}
+\varepsilon_{0, -\bm{q}}
\right)\nonumber\\
&-   
\left(
u_{\bm{q}}
\delta x^\ast_{\bm{q}, \textrm{e}}
+
x^\ast_{\bm{q}, \textrm{e}}
\delta u_{\bm{q}}
\right)\Delta^{\rm{e-e}}_{\bm{q}}
-   
x_{\bm{q}, \textrm{e}}
{\Delta^\ast_{\bm{q}}}^{\rm{e-e}}
\delta u_{\bm{q}}
\nonumber\\
&
- 
\left(
u_{\bm{q}}
\delta x^\ast_{\bm{q}, \textrm{h}}
+
x^\ast_{\bm{q}, \textrm{h}}
\delta u_{\bm{q}}
\right)
\Delta^{\rm{h-h}}_{\bm{q}}
-  
x_{\bm{q}, \textrm{h}}
{\Delta^\ast_{\bm{q}}}^{\rm{h-h}}
\delta u_{\bm{q}}
\nonumber\\
&-
\sum_{S, S_z}
\left[
u_{\bm{q}}
\delta v^\ast_{\bm{q}, S, S_z}
+
v^\ast_{\bm{q}, S, S_z}
\delta u_{\bm{q}}
+\delta_{S, 1}\delta_{S_z, +1}
 v_{\bm{q}, S, -S_z}
 \delta w^\ast_{\bm{q}}
 +\delta_{S, 1}\delta_{S_z, -1}
 v_{\bm{q}, S, -S_z}
 \delta w^\ast_{\bm{q}}
\right.\nonumber\\&\left.
-\frac{1}{2}\delta_{S, 1}\delta_{S_z, 0}
\left(v_{\bm{q}, S, -S_z}+v_{-\bm{q}, S, -S_z}\right)
\delta w^\ast_{\bm{q}}
+\frac{1}{2}\delta_{S, 0}\delta_{S_z, 0}
\left(v_{\bm{q}, S, -S_z}+v_{-\bm{q}, S, -S_z}\right)
\delta w^\ast_{\bm{q}}
\right]
\Delta^{\rm{e-h}}_{\bm{q}}
\nonumber\\
&-
\sum_{S, S_z}
\left[
v_{\bm{q}, S, S_z}
\delta u_{\bm{q}}
+\delta_{S, 1}\delta_{S_z, +1}
  w_{\bm{q}}
 \delta v^\ast_{\bm{q}, S, -S_z}
 +\delta_{S, 1}\delta_{S_z, -1}
w_{\bm{q}} 
 \delta v^\ast_{\bm{q}, S, -S_z}
\right.\nonumber\\&\left.
-\frac{1}{2}\delta_{S, 1}\delta_{S_z, 0}
\left(w_{\bm{q}}+w_{-\bm{q}}\right)
\delta v^\ast_{\bm{q}, S, -S_z}
+\frac{1}{2}\delta_{S, 0}\delta_{S_z, 0}
\left(w_{\bm{q}}+w_{-\bm{q}}\right)
\delta v^\ast_{\bm{q}, S, -S_z}\right]
{\Delta^\ast_{\bm{q}}}^{\rm{e-h}}.
\end{align}
The condition $\delta\langle\Psi|H|\Psi\rangle=0$ leads to the variational equations of $v_{\bm{q},S,S_z}$, $x_{\bm{q},{\rm e(h)}}$, and $w_{\bm{q}}$ shown in the main text.
\end{widetext}

\section{Exciton energy}\label{appendix}

Here, we derive the exciton energy in the present model with the contact-type electron-hole interaction.
The two-body wave function for a $S_z=+1$ exciton reads
\begin{align}
    |\psi_2\rangle
    &=\sum_{\bm{q}}\phi_{\bm{q}}E_{1,+1}^\dag(0,\bm{q})|0\rangle,
\end{align}
where $|0\rangle$ is the vacuum state.
The variational equation with respect to $\phi_{\bm{q}}^*$ given by $\frac{\partial }{\partial \phi_{\bm{q}}^*}\langle \psi_2|H_{\rm e}^0+H_{\rm h}^0+V_{\rm e-h}+B_{\rm X}|\psi_2\rangle=0$ leads to 
\begin{align}
\label{eq:phi}
    \phi_{\bm{q}}(\varepsilon_{{\rm e}, \bm{q}}+\varepsilon_{{\rm h},-\bm{q}}-W)=U_{\rm C}\sum_{\bm{p}}\phi_{\bm{p}}.
\end{align}
Eliminating $\phi_{\bm{q}}$ from Eq.~(\ref{eq:phi}),
one can obtain
\begin{align}
\label{eq:2beq}
    1=U_{\rm C}\sum_{\bm{p}}\frac{1}{q^2/(2M_{\rm r})+B_{\rm X}},
\end{align}
where we have introduced the reduced mass $M_{\rm r}^{-1}=M_{\rm e}^{-1}+M_{\rm h}^{-1}$ and taken $\mu_{\rm e}=\mu_{\rm h}=0$.
Performing the momentum integration in Eq.~(\ref{eq:2beq}),
we obtain
\begin{align}
    \frac{\pi^2}{U_{\rm C}M_{\rm r}}=\Lambda+\sqrt{2M_{\rm r}B_{\rm X}}\tan^{-1}\left(\frac{\Lambda}{\sqrt{2M_{\rm r}B_{\rm X}}}\right),
\end{align}
where $\Lambda$ is the momentum cutoff.
In the limit of $\Lambda\gg \sqrt{2M_{\rm r}B_{\rm X}}$,
we obtain the exciton energy as
\begin{align}
 B_{\rm X}=\frac{1}{2M_{\rm r}}\left(\frac{2\Lambda}{\pi}-\frac{2\pi}{M_{\rm r}U_{\rm C}}\right)^2.   
\end{align}

\end{CJK}
\end{document}